\documentstyle[twocolumn]{mn}
\newcommand{\mincir}{\raise
-2.truept\hbox{\rlap{\hbox{$\sim$}}\raise5.truept
\hbox{$<$}\ }}
\newcommand{\magcir}{\raise
-2.truept\hbox{\rlap{\hbox{$\sim$}}\raise5.truept
\hbox{$>$}\ }}
\newcommand{\minmag}{\raise-2.truept\hbox{\rlap{\hbox{$<$}}\raise
6.truept\hbox
{$>$}\ }}
\newcommand{\be}{\begin{equation}}
\newcommand{\ee}{\end{equation}}
\newcommand{\ba}{\begin{eqnarray}}
\newcommand{\ea}{\end{eqnarray}}
\newcommand{\brr}{\begin{array}}

\newcommand{\err}{\end{array}}
\newcommand{\bc}{\begin{center}}
\newcommand{\ec}{\end{center}}

\input{psfig.sty}
\title
{Predicting the clustering properties of galaxy clusters detectable for
the {\em Planck\/} satellite}
\author[Moscardini et al.]
{
L. Moscardini$^1$, M. Bartelmann$^2$,
S. Matarrese$^3$ and P. Andreani$^4$ \\
$^1$Dipartimento di Astronomia, Universit\`a di
Padova, vicolo dell'Osservatorio 2, I--35122 Padova, Italy\\
$^2$Max-Planck-Institut f\"ur Astrophysik, Karl-Schwarzschild-Strasse
1, D-85748 Garching, Germany \\
$^3$Dipartimento di Fisica G. Galilei, Universit\`{a} di Padova, via
Marzolo 8, I--35131 Padova, Italy\\
$^4$Osservatorio Astronomico di Padova,
vicolo dell'Osservatorio 5, I--35122 Padova, Italy }

\begin{document}

\maketitle

\begin{abstract}
We study the clustering properties of the galaxy clusters detectable
for the {\em Planck\/} satellite due to their thermal
Sunyaev-Zel'dovich effect. We take the past light-cone effect and the
redshift evolution of both the underlying dark matter correlation
function and the cluster bias factor into account. A theoretical
mass-temperature relation allows us to convert the sensitivity limit
of a catalogue into a minimum mass for the dark matter haloes hosting
the clusters. We confirm that the correlation length is an increasing
function of the sensitivity limits defining the survey.  Using the
expected characteristics of the {\em Planck\/} cluster catalogue,
which will be a quite large and unbiased sample, we predict the
two-point correlation function and power spectrum for different
cosmological models. We show that the wide redshift distribution of
the {\em Planck\/} survey, will allow to constrain the cluster
clustering properties up to $z\approx 1$. The dependence of our
results on the main cosmological parameters (the matter density
parameter, the cosmological constant and the normalisation of the
density power-spectrum) is extensively discussed. We find that the
future {\em Planck\/} clustering data place only mild constraints on
the cosmological parameters, because the results depend on the
physical characteristics of the intracluster medium, like the baryon
fraction and the mass-temperature relation. Once the cosmological
model and the Hubble constant are determined, the clustering data will
allow a determination of the baryon fraction with an accuracy of few
per cent.
\end{abstract}


\section{Introduction}

In the standard picture of structure formation based on the
gravitational instability paradigm, clusters of galaxies represent the
largest gravitationally bound systems in the universe. This is the
reason for their cosmological importance. In fact, since the expected
displacements from their primordial positions are much smaller than
their typical separations, clusters retain the imprint of the main
cosmological parameters, which can hopefully be constrained by
studying their properties. In the past this opportunity has been
largely exploited using cluster counts and their redshift evolution
(e.g. Eke et al. 1998; Sadat, Blanchard \& Oubkir 1998; Viana \&
Liddle 1999; Borgani et al. 2001; Verde, Haiman \& Spergel 2001),
obtaining tight constraints on the matter density parameter
$\Omega_{\rm 0m}$ and power-spectrum normalisation $\sigma_8$.

Galaxy clusters are also good tracers of the large-scale structure of
the Universe. Their clustering signal can easily be detected, even
with a relatively small number of objects. In fact, they form in
overdense regions of the cosmological density field and are strongly
biased. A confirmation that the spatial distribution of galaxy
clusters is highly correlated came from the statistical analysis of
the first extended optical surveys (Nichol et al. 1992; Croft et
al. 1997), which however may be affected by the presence of interloper
galaxies. A better identification of galaxy clusters is possible in
the X-ray band. Here their emission due to thermal bremsstrahlung from
the hot intracluster plasma, is more centrally concentrated because it
depends on the square of the baryon density. In the past years the
completion of extended catalogues covering a large fraction of the sky
allowed the computation of the clustering properties of X-ray selected
clusters (Abadi, Lambas \& Muriel 1998; Borgani, Plionis \&
Kolokotronis 1999; Moscardini et al. 2000a; Collins et al. 2001;
Schuecker et al. 2001). The results obtained for different surveys are
fairly different, reflecting a strong dependence on the cluster
selection criteria (limiting flux or luminosity in a given band)
adopted.

It is important to notice that the clustering of galaxy clusters
represents an important test for models of structure formation. As
shown by numerical simulations, cluster evolution can easily be
understood and interpreted, once they are identified as the most
massive dark haloes produced by gravitational collapse. As a
consequence, one can model their clustering properties starting from
the dark halo properties, which can be obtained from an extension of
the standard Press-Schechter formalism (e.g. Mo \& White 1996). The
resulting correlation function depends strongly on cosmology and the
primordial power spectrum. In this paper we will use an improved
theoretical model (Matarrese et al. 1997), paying particular attention
to the redshift evolution of object clustering and to light-cone and
selection effects. A treatment of the redshift-space distortions is
also included. This model has already been applied recently to compare
the existing cluster data (both in the optical and in the X-ray bands)
with model predictions, finding further evidence in favour of
low-density models (Moscardini et al. 2000b; Moscardini, Matarrese \&
Mo 2001).

In this context, a new window is now opening to extend the study of
the cluster distribution in the microwave band, thanks to the thermal
Sunyaev-Zel'dovich (SZ) effect, i.e. the change of the blackbody
spectrum of the cosmic radiation background produced by Compton
scattering off the hot intracluster medium. Observational data of some
individual galaxy clusters have already been obtained, allowing a
better morphological study of individual objects. However, at present
well-defined surveys required for statistical studies are not
available yet, even though several ground-based experiments with this
objective are being planned for the near future (e.g. BOLOCAM, AMIBA,
CBI, AMI, etc.).

The goal of a full-sky survey of SZ clusters will be reached only with
space missions. In particular, the {\em Planck\/} satellite, which is
scheduled for launch in 2007, will detect of order $10^4$ clusters
thanks to the optimal choice of filter bands around the frequencies
relevant for the SZ effect. This large number of clusters, even if
obtained with an angular resolution lower than that of the
above-mentioned ground-based bolometer arrays and interferometers,
will allow an accurate measurement of the cluster correlation function
and constraints on its redshift evolution.

This paper reports on an extended study of the clustering properties
of galaxy clusters detectable for the {\em Planck\/} satellite. It is
worth mentioning that an estimate of the contribution from the
correlation among clusters to the angular power spectrum of the cosmic
microwave background radiation anisotropy due to fluctuations of the
SZ effect has been obtained by Komatsu \& Kitayama (1999) using an
approach similar to the one applied here (see also Refregier et
al. 2000).

The plan of the paper is as follows.
In Section 2 we discuss the general characteristics of the objects
detectable for {\em Planck\/}. Section 3 reviews the theoretical model
for the correlation function of SZ galaxy clusters in the framework of
different cosmological models. In Section 4 we discuss the
implications of the previous results and the possibility of
constraining the main cosmological parameters using the future {\em
Planck\/} observations. Section 5 shows the dependence of the
clustering predictions on the characteristics of the intracluster
medium, like the baryon fraction and the mass-temperature
relation. Conclusions are drawn in Section 6.

\section{Galaxy clusters detectable for {\em Planck\/} through thermal
Sunyaev-Zel'dovich emission }

Galaxy clusters can be detected in the microwave regime due to their
thermal Sunyaev-Zel'dovich effect. Hot electrons in the intracluster
medium Compton-scatter the cold photons of the microwave background
and re-distribute them towards higher frequencies. The result is a
temperature decrement below 218~GHz, and an increment above.

The SZ effect in direction $\vec{\theta}$ is quantified by the
Compton-$y$ parameter,
\be
y(\vec{\theta})= {{k\sigma_{_T}}\over{m_e c^2}} \int dl
\ n_e(\vec{\theta},l)\ T(\vec{\theta},l)
\ ,
\label{eq:y_theta}
\ee
where $T$ is the electron temperature, $n_e$ the three-dimensional
thermal electron density (hereafter assumed to follow a King profile),
and $\sigma_{_T}$ is the Thomson scattering cross section.

Assuming an isothermal distribution of gas and neglecting the
background noise (see the following discussion), the total SZ signal
received from a cluster is simply the integral over the solid angle
covered, i.e.
\be
Y\equiv \int d^2\ \vec{\theta}\ y(\vec{\theta})= {{kT}\over {m_e c^2}}
{\sigma_{_T}\over D_d^2} N_e \ ,
\ee
where $D_d$ is the angular-diameter distance to the cluster, which of
course depends on cosmology. The total number of electrons in the
cluster $N_e$ can be assumed to be proportional to the virial mass
$M$,
\be
N_e={{1+f_H}\over 2} f_b {M\over m_p}\ ,
\ee
where $m_p$ is the proton mass, $f_b$ is the baryon fraction of the
cluster mass and $f_H$ is the hydrogen fraction of the baryonic mass
(here assumed as 76 per cent). Assuming an isothermal gas distribution
in virial equilibrium, it is possible to relate the cluster
temperature (in keV) to its mass,
\be
T = 6.03 {\left(M\over {10^{15} h^{-1} M_\odot}\right)}^{2/3}
E^{2/3}(z)
\left[{\Delta_{\rm vir}(z) \over {178}}\right]^{1/3} \ .
\label{eq:t-m}
\ee
The quantity $\Delta_{\rm vir}(z)$ is the mean density (in units of
the critical density at redshift $z$) of the virialised halo computed
from the spherical collapse model; $E(z)$ is the Hubble constant at
redshift $z$ in units of its present value, $E(z)
\equiv H(z)/H_0$. The normalisation of the mass-temperature relation
is taken from the analysis of hydrodynamical simulations by Mathiesen
\& Evrard (2001). Using this set of relations, the quantity $Y$
depends on the cosmological model, and on cluster mass and redshift
only, i.e. $Y\equiv Y(M,z)$.

For modelling the cluster population detectable for {\em Planck\/} we
have to take into account the characteristics of the satellite
detector (e.g. Haehnelt 1997; Bartlett 2000; Bartelmann 2001). Its
relatively poor angular resolution will not allow most of the cluster
population to be resolved (e.g. Aghanim et al. 1997; Hobson et
al. 1998). Unresolved clusters below the detection limit will produce
a Compton-$y$ background $y_{\rm bg}$. Neglecting background
correlations, the average background fluctuation level can be
estimated as
\be
 (\Delta y_{\rm bg})^2=\int dz \left| {{dV}\over{dz}} \right|
 (1+z)^3 \int dM \ \overline{n}(M,z) \ Y^2(M,z)
\label{eq:background}
\ee
(e.g. Bartelmann 2001). In the previous expression, $dV$ is the cosmic
volume per unit redshift and unit solid angle, and $\overline{n}(M,z)$
is the cluster mass function which can be estimated using the
Press-Schechter approach or more recent extensions, see below.

Moreover, the original cluster SZ signal will be convolved by the {\em
Planck\/} beam profile $w(\vec{\theta})$, which depends on the
observing frequency. We assume for our purposes that cluster selection
will be mainly performed at the highest frequencies and approximate
the beam profile with a Gaussian with r.m.s. $\sigma_w=5$ arcmin.

We assume that a cluster is detectable for {\em Planck\/} if its
integrated, beam-convolved Compton-$y$ parameter is larger than a
given sensitivity limit, i.e. $\overline{Y} \ge
\overline{Y}_{\rm min}$, where
\be
\overline{Y}\equiv {1\over {2\pi\sigma_w^2}}
\int d^2\theta \int d^2 \theta^{\prime} \ y(\theta^{\prime})\ \exp\left[
-{{(\theta-\theta^{'})^2}\over{2\sigma_w^2}}
\right] \ .
\ee
The external integral covers the solid angle where the signal is large
enough to be detected. Using its nominal temperature sensitivity and
the solid angle of the beam, the sensitivity limit for {\em Planck\/}
can conservatively be set to $\overline{Y}_{\rm min}=3\times 10^{-4}$
arcmin$^2$ (Bartelmann 2001).

\section{The clustering model}

Our model for predicting the clustering properties of the SZ galaxy
clusters derives from a method already applied to the study of
clusters detected in the X-ray and optical bands (Moscardini et
al. 2000a,b; Moscardini, Matarrese \& Mo 2001). We will give a brief
description of the technique only and refer to the original papers for
a more detailed discussion.

Matarrese et al. (1997; see also Moscardini et al. 1998; Yamamoto \&
Suto 1999; Suto et al. 2000; Hamana et al. 2001) developed an
algorithm for describing the clustering on our past light-cone taking
into account both the non-linear dynamics of the dark matter
distribution and the redshift evolution of the bias factor. The final
expression for the observed spatial correlation function $\xi_{\rm
obs}$ in a given redshift interval ${\cal Z}$ is
\be
\xi_{\rm obs}(r) = { \int_{\cal Z}
d z_1 d z_2 {\overline{\cal N}}(z_1) {\overline{\cal N}}(z_2)
~\xi_{\rm obj}(r;z_1,z_2) \over \bigl[ \int_{\cal Z} d z_1
{\overline{\cal
N}}(z_1) \bigr]^2 } \;,
\label{eq:xifund}
\ee
where ${\overline{\cal N}}(z)\equiv {\cal N}(z)/r(z)$, ${\cal N}(z)$
is the actual redshift distribution of the catalogue and $r(z)$
describes the relation between the comoving radial coordinate and the
redshift.  In (\ref{eq:xifund}), $\xi_{\rm obj}(r,z_1,z_2)$ represents
the correlation function of pairs of objects at redshifts $z_1$ and
$z_2$ with comoving separation $r$, which can be safely approximated
as $\xi_{\rm obj}(r,z_1,z_2) \approx b_{\rm eff}(z_1) b_{\rm eff}(z_2)
\xi_{\rm m}(r,z_{\rm ave})$. Here $\xi_{\rm m}$ is the dark matter
covariance function and $z_{\rm ave}$ is a suitably defined
intermediate redshift (see Porciani 1997 for a discussion of possible
choices).

A fundamental role in the previous equation is played by the effective
bias $b_{\rm eff}$, which can be expressed as a weighted average of
the `monochromatic' bias factor $b(M,z)$ of objects with some given
intrinsic property $M$ (like mass, luminosity, etc):
\be
b_{\rm eff}(z) \equiv {\cal N}(z)^{-1} \int_{\cal M} d\ln M' ~b(M',z)
~{\cal N}(z,M')\, ,
\label{eq:b_eff}
\ee
where ${\cal N}(z,M)$ is the number of objects actually present in the
catalogue with redshift within $dz$ of $z$ and mass within $dM$ of
$M$, whose integral over $\ln M$ is ${\cal N}(z)$.

%--------------------------------------------------------
\begin{figure}
\centering
\psfig{figure=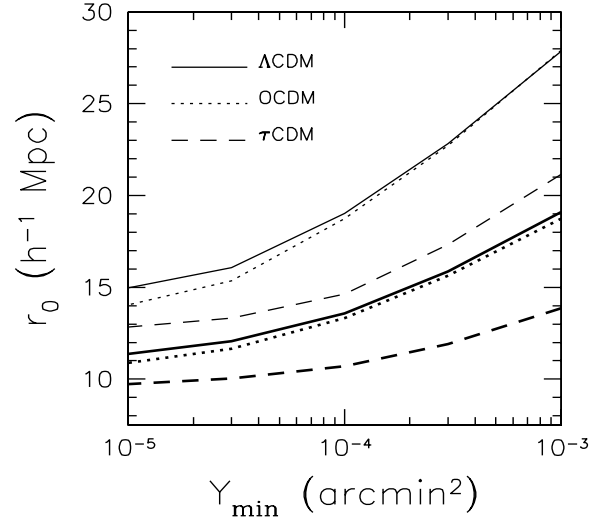,width=8.cm,height=8.cm,angle=0}
\caption{
The dependence of the correlation length $r_0$ on the sensitivity
limit $\overline{Y}_{\rm min}$ (in arcmin$^2$) is shown for $\Lambda
CDM$ (solid lines), OCDM (dotted lines) and $\tau CDM$ models (dashed
lines). Heavy lower and light upper lines refer cluster samples with
$z<0.3$ and $z>0.3$, respectively.}
\label{fi:y_min}
\end{figure}
%--------------------------------------------------------
%

In the case of galaxy clusters, it is possible to fully characterise
their properties by the mass $M$ of their hosting dark matter haloes
at each redshift $z$. This is consistent with the hierarchical model
of structure formation where clusters are expected to form by merging
of smaller mass units. The comoving mass function $\bar n(z,M)$ of
dark matter haloes, required both in eq. (\ref{eq:background}) and in
the computation of ${\cal N}(z,M)$, can be estimated using some
extension of the Press-Schechter formalism. Moreover, we can adopt the
Mo \& White (1996) relation (or improvements thereof; see Sheth \&
Tormen 1999) for the monochromatic bias required in
eq. (\ref{eq:b_eff}). More precisely, we adopt the analytic relations
obtained by Sheth \& Tormen (1999; see also Sheth, Mo \& Tormen 2001)
which are very well reproduced by the results of high-resolution
$N$-body simulations (Jenkins et al. 2001).

For incorporating the fully non-linear regime, we use the fitting
formula by Peacock \& Dodds (1996) which allows the analytic
computation of the redshift evolution of the dark matter covariance
function $\xi_{\rm m}$. We also include the effects of redshift-space
distortions using linear theory and the distant-observer approximation
(Kaiser 1987).

Finally, in order to predict the abundance and clustering of galaxy
clusters detected through their thermal SZ emission, we need to relate
the expected characteristics of the samples (which in the case of SZ
detected objects can be expressed in terms of the sensitivity limit
$\overline{Y}_{\rm min}$) to a corresponding halo mass at each
redshift. Such a relation is provided by the mass-temperature relation
(eq. \ref{eq:t-m}).

Changing the sensitivity limits, the characteristics of the detected
cluster population change. In deep catalogues, faint objects
(corresponding to low-mass haloes) can be detected. Due to the strong
mass dependence of the monochromatic bias, a different clustering
amplitude is expected. In Fig.~\ref{fi:y_min} we show the dependence
of the correlation length $r_0$ (defined as the scale where $\xi_{\rm
obs}$ is unity) on the limits used to define the SZ surveys, i.e. the
limiting sensitivity $\overline{Y}_{\rm min}$. The results are shown
for galaxy clusters having $z<0.3$ (heavy lower lines) and $z>0.3$
(light upper lines) because at this redshift the expected {\em
Planck\/} sample is roughly divided into two subsamples with a similar
number of objects. We consider three different cosmological models,
whose parameters are listed in Table \ref{t:models}. We find that
$r_0$ is an increasing function of the limits defining a cluster
survey: the larger $\overline{Y}_{\rm min}$ is, the more clustered the
objects are, independent of the cosmological model. It is interesting
to note that a similar trend was also predicted for galaxy clusters
detected in the X-ray (Moscardini et al. 2000b) and optical bands
(Moscardini et al. 2001).

\section{Cosmological implications}

We now discuss how the clustering properties of the SZ clusters depend
on the cosmological parameters. We set the SZ sensitivity limit to
that expected for the {\em Planck\/} satellite,
i.e. $\overline{Y}_{\rm min}=3\times 10^{-4}$ arcmin$^2$. We will
consider in detail three different cosmological models. They have a
cold dark matter (CDM) power spectrum, a primordial power spectral
index $n=1$, and a shape parameter $\Gamma=0.21$ in agreement with an
extended set of observational data (e.g. Peacock \& Dodds
1996). Moreover, they have the power-spectrum normalisation obtained
by Viana \& Liddle (1999) to reproduce the local cluster
abundance. Notice that this normalisation has been challenged by very
recent analyses, which, even if based on quite different approaches,
seem to converge to substantially smaller values for $\sigma_8$
(Reiprich \& B\"ohringer 2002; Viana, Nichol \& Liddle 2002; Seljak
2001).  Combined analyses of data coming from the 2dF
Galaxy Redshift Survey and from measurements of the cosmic microwave
background anisotropies seem to corroborate these low values of the
power spectrum normalisation (Lahav et al. 2002).  We will discuss
later the effect of changing the assumed value for $\sigma_8$. The
models correspond to different geometries. Specifically, we consider a
flat model with matter density parameter $\Omega_{\rm 0m}=0.3$ and
cosmological constant (hereafter $\Lambda$CDM); an open model with
$\Omega_{\rm 0m}=0.3$ and vanishing cosmological constant (OCDM); and
an Einstein-de Sitter model ($\tau$CDM). The two low-density models
have a local Hubble constant $h\equiv H_0/(100 {\rm km/s/Mpc})=0.7$,
while we assume $h=0.5$ for the $\tau$CDM model. The baryon fraction
is set to $f_b=0.075 h^{-3/2}$, as suggested by the analysis of X-ray
data made by Mohr, Mathiesen \& Evrard (1999). Table
\ref{t:models} summarises the model parameters.

\begin{table}
\centering
\caption[]{The parameters of the cosmological models. Column 2: the present
matter density parameter $\Omega_{\rm 0m}$; Column 3: the present
cosmological constant contribution to the density $\Omega_{0\Lambda}$;
Column 4: the primordial spectral index $n$; Column 5: the local
Hubble parameter $h$; Column 6: the shape parameter $\Gamma$; Column
7: the spectrum normalisation $\sigma_8$; Column 8: the baryon
fraction $f_b$.}
\tabcolsep 4pt
\begin{tabular}{lccccccc} \\ \\ \hline \hline
Model & $\Omega_{\rm 0m}$ & $\Omega_{0\Lambda}$ & $n$ & $h$ &
$\Gamma$ & $\sigma_8$ & $f_b$ \\ \hline
$\Lambda$CDM & 0.3 & 0.7 & 1.0 & 0.7 & 0.21 & 0.99 & 0.128 \\
OCDM         & 0.3 & 0.0 & 1.0 & 0.7 & 0.21 & 0.84 & 0.128 \\
$\tau$CDM    & 1.0 & 0.0 & 1.0 & 0.5 & 0.21 & 0.56 & 0.212 \\
\hline
\end{tabular}
\label{t:models}
\end{table}

\begin{figure*}
\centering
\psfig{figure=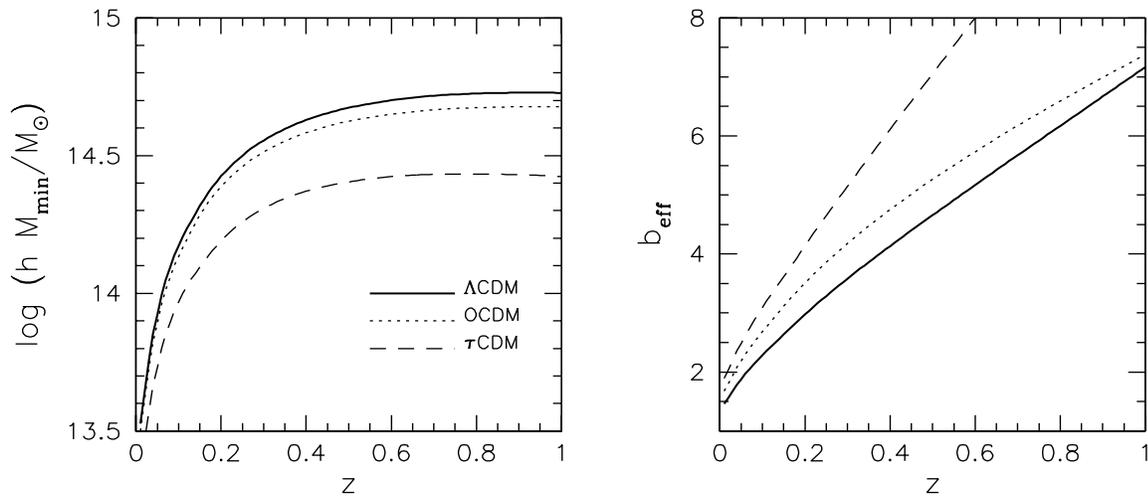,width=16.cm,height=8.cm,angle=0}
\caption{The minimum mass (left panel) and
the effective bias factor (right panel) for clusters detectable for
{\em Planck\/} are plotted as functions of redshift. Different line
types refer to different cosmological models: $\Lambda CDM$ (solid
lines), OCDM (dotted lines), $\tau CDM$ (dashed lines). The model
parameters are summarised in Table
\ref{t:models}.}
\label{fi:mmin_bias}
\end{figure*}

In the left panel of Figure \ref{fi:mmin_bias} we show how the
properties of the cluster population detectable for {\em Planck\/}
change as a function of redshift. Locally objects with a relatively
small mass will be detectable, of the order of galaxy groups: in fact
$M_{\rm min}\la 10^{14}\ h^{-1}\ M_\odot$ for $z\la 0.1$, independent
of the cosmological model. At higher redshifts, the minimum mass
increases: at $z\approx 0.5$ the {\em Planck\/} catalogue will include
clusters with mass larger than $M_{\rm min}\approx 10^{14.6}\ h^{-1}\
M_\odot$ for low-density models and $M_{\rm min}\approx 10^{14.3}\
h^{-1}\ M_\odot$ for the Einstein-de Sitter model. Note that the
differences between $\Lambda$CDM and OCDM are almost negligible. In
fact the SZ clusters detected by {\em Planck\/} have relatively small
redshifts, consequently the presence of the cosmological constant has
small impact on the distance relations and the mass-temperature
relation.

The right panel of Figure \ref{fi:mmin_bias} presents the redshift
dependence of the effective bias factor $b_{\rm eff}$. As expected,
the bias is a strongly increasing function of $z$. This is due to two
different reasons. First, the bias is itself an increasing function of
mass, and we found an increase of $M_{\rm min}$ with redshift; second,
even fixing the halo mass, the (squared) bias is inversely
proportional to the clustering of the underlying dark matter
distribution which is growing with time. This also explains the
largest values of $b_{\rm eff}$ obtained for the $\tau$CDM model: in
fact the growth factor is larger in an Einstein-de Sitter cosmology
than in low-density models.

%--------------------------------------------------------
%--------------------------------------------------------
%

\begin{figure}
\centering
\psfig{figure=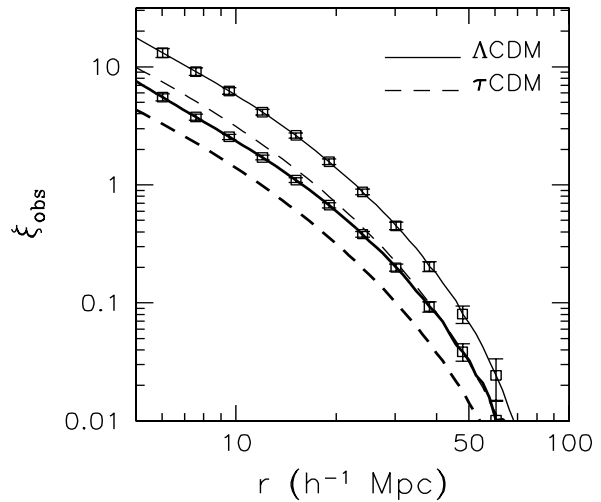,height=8.cm,width=8cm,angle=0}
\caption{The spatial correlation function $\xi_{\rm obs}(r)$ is shown
for clusters detectable for {\em Planck\/} with $z<0.3$ (lower heavy
lines) and $z>0.3$ (upper light lines). The results refer to $\Lambda
CDM$ and $\tau CDM$ models (solid and dashed lines,
respectively). Results for OCDM are not shown because they are almost
indistinguishable from the $\Lambda CDM$ model. Error bars (shown only
for $\Lambda CDM$) are 1-$\sigma$ bootstrap estimates. }
\label{fi:cosmo_csi}
\end{figure}

Figure \ref{fi:cosmo_csi} presents the predicted two-point correlation
function $\xi_{\rm obs}$ computed for cluster samples having $z<0.3$
and $z>0.3$ (lower and upper lines, respectively). In the plot
`mock-observational' 1-$\sigma$ error bars are given (only for the
$\Lambda CDM$ model, for clarity). They are obtained by bootstrap
resampling the number of expected pairs in each separation bin (Mo,
Jing \& B\"orner 1992). $\xi_{\rm obs}$ for the OCDM model is almost
indistinguishable from the $\Lambda CDM$ case. This confirms that the
clustering of SZ clusters is almost insensitive to the cosmological
constant, as already found for objects detected in the X-ray and
optical bands (see also the discussion of the following Figure
\ref{fi:r0_omega}). The predicted correlation function for the
$\tau$CDM model is lower by approximately a factor of two, both for
$z<0.3$ and $z>0.3$. The correlation lengths $r_0$ (still defined as
the scales where $\xi_{\rm obs}$ is unity) for the low-redshift
catalogues are $15.9\pm 0.4$, $15.7\pm 0.6$ and $11.9\pm 0.2$ $h^{-1}$
Mpc for the $\Lambda$CDM, OCDM and $\tau$CDM models, respectively. As
expected from the behaviour of the bias factor displayed in Figure
\ref{fi:mmin_bias}, we find that the clusters are more clustered at
higher redshift: the corresponding values of $r_0$ for the sample with
$z>0.3$ are $22.8\pm 0.7$, $22.7\pm 1.2$ and $17.3\pm 1.1$ $h^{-1}$
Mpc, for $\Lambda$CDM, OCDM and $\tau$CDM models, respectively.
Again, 1-$\sigma$ bootstrapping error estimates are given.

%
%--------------------------------------------------------
%--------------------------------------------------------
%

\begin{figure}
\centering
\psfig{figure=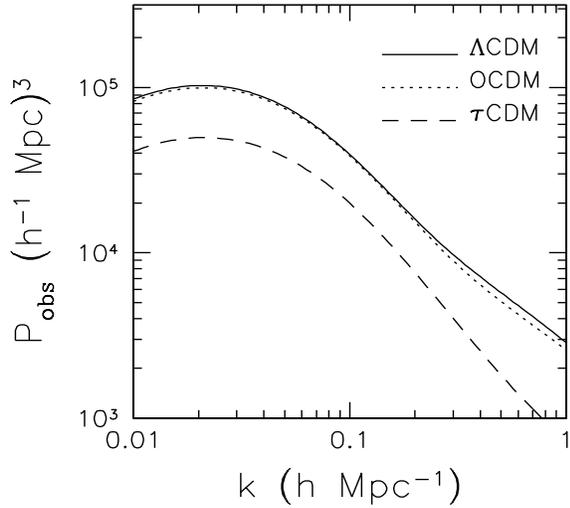,height=8.cm,width=8cm,angle=0}
\caption{Theoretical predictions for the power-spectrum $P_{\rm obs}(k)$
are displayed for the {\em Planck\/} sample of SZ galaxy clusters.
Results are presented for $\Lambda CDM$ (solid line), OCDM (dotted
line) and $\tau CDM$ models (dashed line) and refer to clusters with
$z<0.3$ only.}
\label{fi:cosmo_pk}
\end{figure}

The differences expected between low- and high-density models are
confirmed by the power-spectrum $P_{\rm obs}(k)$, which is shown in
Fig.~\ref{fi:cosmo_pk}. Results are given for the cluster sample with
$z<0.3$ only. Again the differences between open and flat models with
$\Omega_{\rm 0m}=0.3$ are negligible, while the power-spectrum for the
$\tau CDM$ model is approximately a factor of two lower.

%
%--------------------------------------------------------
%--------------------------------------------------------
%

\begin{figure}
\centering
\psfig{figure=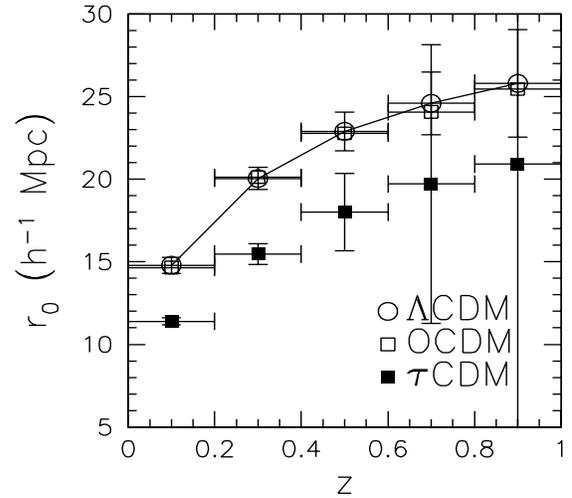,width=8.cm,height=8.cm,angle=0}
\caption{The redshift evolution of the correlation length $r_0$ is
shown for $\Lambda CDM$ (open circles), OCDM (open squares) and $\tau
CDM$ models (filled squares). Results are obtained dividing the
expected cluster samples in redshift bins with size $\Delta
z=0.2$. Error bars (for clarity only shown for the $\Lambda CDM$ and
$\tau CDM$ models) are 1-$\sigma$ bootstrap estimates. }
\label{fi:evol_z}
\end{figure}

In order to better quantify the redshift evolution of clustering, we
show in Fig.~\ref{fi:evol_z} the values of the correlation length
computed using redshift bins of size $\Delta z=0.2$ for the three
cosmological models. From $z\approx 0$ to $z\approx 1$, $r_0$ changes
by a factor of almost two. Locally the SZ clusters detected by {\em
Planck\/} are expected to have a correlation function only slightly
larger than (local) galaxy groups or poor (optical) clusters. The
predicted clustering signal at high redshift is quite large (albeit
more uncertain due to the smaller number of objects) and can certainly
be measured, at least in the case of low-density models. Note that,
given its very large area, the {\em Planck\/} survey will be one of
the best opportunities to measure the redshift evolution of cluster
clustering.  An alternative possibility (but in the X-ray band) will
be the survey proposed for the XMM/Newton satellite (Refregier,
Valtchanov \& Pierre 2001). In this case the expected limiting flux
will be approximately $S_{\rm lim}=5 \times 10^{-15}$ erg cm$^{-2}$
s$^{-1}$ in the 0.5--2 keV band, but over a limited area of 64 square
degrees.

%--------------------------------------------------------
%--------------------------------------------------------
%

So far, we considered models with fixed cosmological parameters. Now
it is interesting to discuss the possibility of
constraining these parameters using the clustering properties of
{\em Planck\/} clusters. We start by considering the changes of the
correlation length $r_0$ when the matter density parameter
$\Omega_{\rm 0m}$ is varied. The results are presented in Figure
\ref{fi:r0_omega} for galaxy clusters with $z<0.3$ and $z>0.3$ (heavy and
light lines, respectively). All models shown here have the spectrum
normalisation required to reproduce the local cluster abundance. We
use again the relation found by Viana \& Liddle (1999), namely
$\sigma_8=0.56 \Omega_{\rm 0m}^{-C}$, where $C=0.34$ and $C=0.47$ for
open and flat models, respectively. Different choices for the
normalisation will be discussed later. Primordial spectral index,
shape parameter, Hubble parameter and baryon fraction are taken from
Tab.~\ref{t:models}. We find that varying $\Omega_{\rm 0m}$ from $0.2$
to $1.0$ changes $r_0$ by 50 per cent for clusters with $z<0.3$, and
30 per cent for clusters with $z>0.3$. More precisely, the predicted
correlation length decreases as $\Omega_{\rm 0m}$ increases.
Moreover, we notice that the cosmological constant increases the
correlation length by less than 5\%. In particular, at high redshifts
the curves for flat and open models are almost identical. This is due
to the relatively large value of the minimum mass of the detected
clusters which implies the inclusion of objects with very similar
properties. A similar result, i.e. the impossibility of constraining
the presence of the cosmological constant using the clustering of
galaxy clusters, was also found analysing cluster data in the optical
and X-ray bands (Moscardini et al. 2000b; Moscardini, Matarrese \& Mo
2001).

%--------------------------------------------------------
\begin{figure}
\centering
\psfig{figure=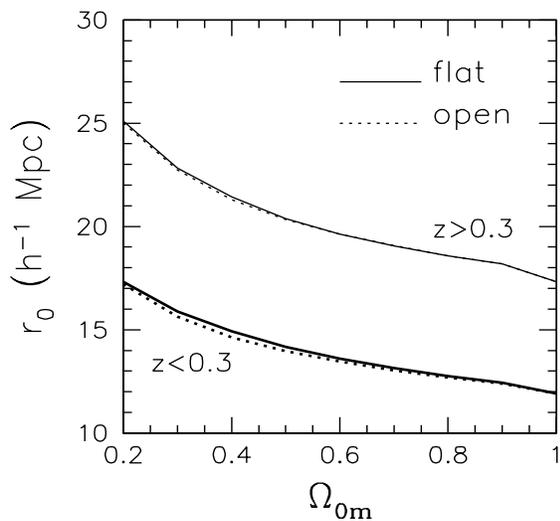,width=8.cm,height=8.cm,angle=0}
\caption{The predicted correlation length $r_0$ is plotted as
a function of the present matter density parameter $\Omega_{\rm 0m}$.
Results are presented for CDM models with $\Gamma=0.21$ and $\sigma_8$
chosen to reproduce the local cluster abundance (Viana \& Liddle
1999). Solid lines: flat cosmological models (i.e. with non-zero
cosmological constant); dotted lines: open models with vanishing
$\Omega_{\rm 0\Lambda}$. Heavy lower and light upper lines refer to
galaxy clusters with $z<0.3$ and $z>0.3$, respectively.}
\label{fi:r0_omega}
\end{figure}
%--------------------------------------------------------
%

\begin{figure}
\centering
\psfig{figure=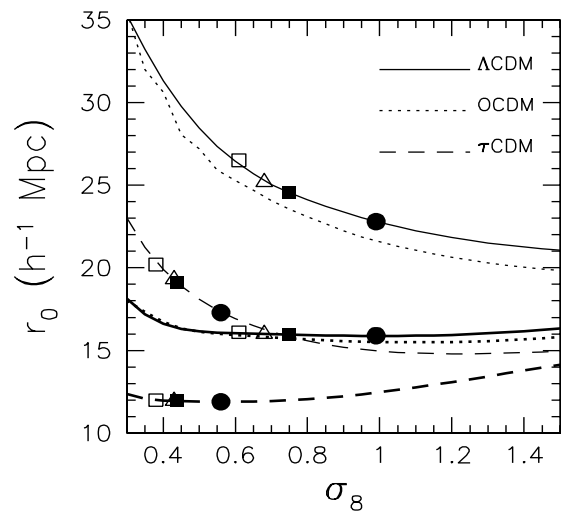,width=8.cm,height=8.cm,angle=0}
\caption{The predicted correlation length $r_0$ is shown as a function
of the spectrum normalisation $\sigma_8$. Heavy lower and light upper
lines refer to clusters with $z<0.3$ and $z>0.3$,
respectively. Results are presented for $\Lambda CDM$ (solid lines),
OCDM (dotted lines) and $\tau CDM$ models (dashed lines). Values of
$r_0$ correspond to different estimates of $\sigma_8$ obtained by
Viana \& Liddle (1999), Seljak (2001), Reiprich \& B\"ohringer (2002)
and Viana, Nichol \& Liddle (2002) are shown by solid circles, solid
squares, open triangles and open squares, respectively.}
\label{fi:r0_dep_new}
\end{figure}

In Figure \ref{fi:r0_dep_new} we show how the correlation length $r_0$
changes when the spectrum normalisation $\sigma_8$ is varied. Again,
results are shown for galaxy clusters with $z<0.3$ and $z>0.3$ (heavy
and light lines, respectively). In the low-redshift bin, $r_0$ shows a
small dependence on $\sigma_8$ and no significant differences are
found using the Viana \& Liddle (1999) normalisation compared to the
more recent values obtained by Reiprich \& B\"ohringer (2002), Seljak
(2001) and Viana, Nichol \& Liddle (2002). However, for clusters with
$z>0.3$ the variation is expected to be quite large. The predicted
correlation length for low-density models changes from $r_0\approx 35$
to $r_0\approx 20 h^{-1}$ Mpc as $\sigma_8$ is increased from 0.3 to
1.5. For the $\tau$CDM model the change is more limited: $r_0\approx
23 h^{-1}$ Mpc for $\sigma_8=0.3$ and $r_0\approx 14-15 h^{-1}$ Mpc
for $0.7\la \sigma_8<1.5$. In particular, when the normalisation is
changed from the largest value (Viana \& Liddle 1999) to the lowest
one (Viana et al. 2002), $r_0$ increases by roughly 20 per cent.

We also checked the effect on cluster clustering when only the Hubble
parameter $h$ is varied, keeping all other parameters fixed as in
Tab.~\ref{t:models} (the results are not shown in the Figure). Varying
$h$ from $0.4$ to $1$, the correlation length changes at most by 10\%,
independent of the cosmological model and redshift bin. This variation
is certainly too small to allow any constraints on the Hubble
constant.

%--------------------------------------------------------
%--------------------------------------------------------
%

\section{Dependence on the properties of the intracluster medium}

The SZ signal of galaxy clusters strongly depends on the properties of
the intracluster medium. In fact the SZ signal reflects the
distribution of hot gas in the clusters, as shown by
eq. (\ref{eq:y_theta}). So far, we employed a simple model: a baryon
fraction fixed by the X-ray data of Mohr et al. (1999), a King profile
for the gas distribution and a virial scaling relation between mass
and temperature. In the following subsections we will study how
sensitive the previous results are to these assumptions. Moreover, we
will discuss if the clustering of SZ galaxy clusters can be used to
directly constrain the properties of the intracluster medium.

\subsection{Baryon fraction}

%--------------------------------------------------------
\begin{figure*}
\centering
\psfig{figure=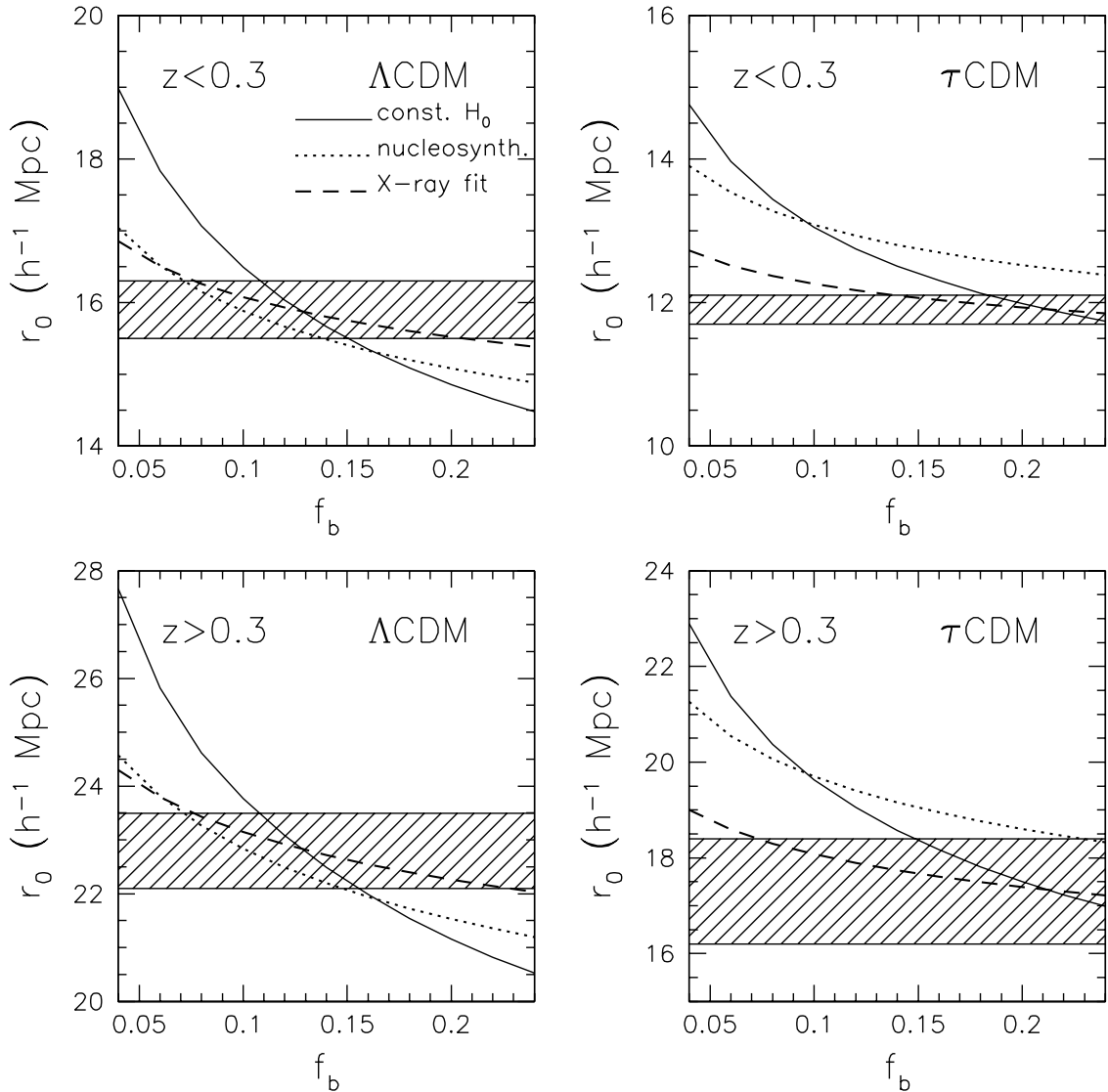,width=16.cm,height=16.cm,angle=0}
\caption{The predicted correlation length $r_0$ of galaxy clusters
with $z<0.3$ (upper panels) and $z>0.3$ (lower panels) is plotted as a
function of the baryon fraction $f_b$. Results are presented for the
$\Lambda$CDM (left panels) and $\tau$CDM models (right panels).
Different lines refer to different assumptions on the relation between
the baryon fraction and the Hubble parameter $h$: constant $h$, equal
to 0.7 and 0.5 for $\Lambda CDM$ and $\tau CDM$ models, respectively
(solid lines); $f_b=0.024/\Omega_{\rm 0m} h^2$ as predicted from
primordial nucleosynthesis (dotted lines); $f_b=0.075 h^{-3/2}$ as
derived from X-ray data (dashed lines). The dashed regions show the
1-$\sigma$ ranges for $r_0$ as obtained using the parameters from
Tab.~\ref{t:models}.}
\label{fi:fbar}
\end{figure*}
%--------------------------------------------------------
%

In Figure \ref{fi:fbar} we show the dependence of the predicted
correlation length $r_0$ on the baryon fraction $f_b$.  We consider
cluster samples with $z<0.3$ and $z>0.3$ (upper and lower panels,
respectively), and two different cosmological models: $\Lambda CDM$
and $\tau CDM$ (left and right panels, respectively). We do not
present results for OCDM because of their similarity with the $\Lambda
CDM$ model. Three different relations between $f_b$ and the present
Hubble parameter $h$ are assumed. First, we assume a perfect knowledge
of the value of the Hubble constant: $h=0.7$ and $h=0.5$ for $\Lambda
CDM$ and $\tau CDM$, respectively. Even if the second value appears to
be too low when compared to recent observational estimates
(e.g. Freedman et al. 2001), it is required in the case of an
Einstein-de Sitter model to avoid a strong conflict between the age of
the Universe and that of globular clusters. Second, we vary $h$ in
agreement with the constraints from primordial nucleosynthesis:
$f_b=0.024/\Omega_{\rm 0m} h^2$ (see e.g. the review by Schramm
1998). Finally we adopt again the relation given by Mohr et al. (1999)
from the analysis of the X-ray data of 45 galaxy clusters. Notice
that, given $f_b$, the values of $h$ corresponding to the previous
relations can vary quite substantially. Consequently, the predicted
scatter of $r_0$ indicates how well $f_b$ can be constrained with
these clustering data.

In general we find small changes when we adopt the relations
from primordial nucleosynthesis and X-ray data: the predicted
values of $r_0$ differ only by 0.5-1 $h^{-1}$ Mpc for $0.05 \la f_b
\la 0.25$, independent of the cosmological model. Thus, even with a
well-determined correlation length, as expected with the {\em
Planck\/} survey, only a rough estimate of the baryon fraction will be
possible.  The situation changes if we assume independent knowledge of
the Hubble parameter $h$. Considering the expected 1-$\sigma$ error
bars on $r_0$ for the low-redshift dataset, $f_b$ can be estimated
from clustering data with typical uncertainties smaller than 0.02-0.03
for the $\Lambda CDM$ model and 0.05 for the $\tau CDM$ model. At
higher redshifts ($z>0.3$), one would expect a stronger dependence of
the correlation length on the baryon fraction. In fact, if the value
of $f_b$ is increased, clusters with lower mass are included in the
{\em Planck\/} catalogue. Since the mass function is so steep, almost
all clusters will have masses near the lower mass limit, which
essentially determines the bias factor and consequently the
correlation length.  This is confirmed by our results: the change of
$r_0$ with $f_b$ at high redshift is quite substantial (about 40 per
cent across the plots). However, the expected 1-$\sigma$ error bars in
the determinations of the correlation length increases in this
redshift bin and the possibility to constrain $f_b$ remains roughly
similar to the $z<0.3$ dataset.

\subsection{Dependence on the parameters of the mass-temperature relation}

The results obtained so far were based on a fixed mass-temperature
relation (eq. \ref{eq:t-m}). The underlying assumption is an
isothermal distribution of gas in virial equilibrium, which is in
discrete agreement both with numerical simulations (e.g. Bryan \&
Norman 1998; Frenk et al. 1999; Mathiesen \& Evrard 2001) and
observational estimates (Xu, Jin \& Wu 2001; Finoguenov, Reiprich \&
B\"ohringer 2002). However the parameters in the relation are not well
known. For example, different groups obtained different numerical
values for the proportionality constant. Moreover, the presence of
radiative cooling in cluster simulations changes both the
normalisation and the slope of the mass-temperature relation
(e.g. Muanwong et al. 2001). Finally also the redshift evolution of
the relation, which cannot be constrained by present observational
data, can be modified by different mechanisms, such as for instance
the presence of an entropy floor (e.g. Tozzi \& Norman 2001).

For testing the robustness of our previous results we will now study
the variation of the correlation length $r_0$ in response to changes
of the parameters in the mass-temperature relation. We modify
eq. (\ref{eq:t-m}) as follows:
\be
T = \alpha_{_1} {\left(M\over {10^{15} h^{-1}
M_\odot}\right)}^{\alpha_{_2}} (1+z)^{\alpha_{_3}} E^{2/3}(z)
\left[{\Delta_{\rm vir}(z) \over {178}}\right]^{1/3} \ .
\label{eq:t-m_mod}
\ee
Here the parameters $\alpha_{_1}$, $\alpha_{_2}$ and $\alpha_{_3}$
represent the normalisation, the slope of the relation and the
exponent of the redshift dependence, respectively. The results of the
previous section correspond to
$(\alpha_{_1},\alpha_{_2},\alpha_{_3})=(6.03,2/3,0)$. The cosmological
model is fixed to the $\Lambda CDM$ model (see Table
\ref{t:models}).

%--------------------------------------------------------
\begin{figure*}
\centering
\psfig{figure=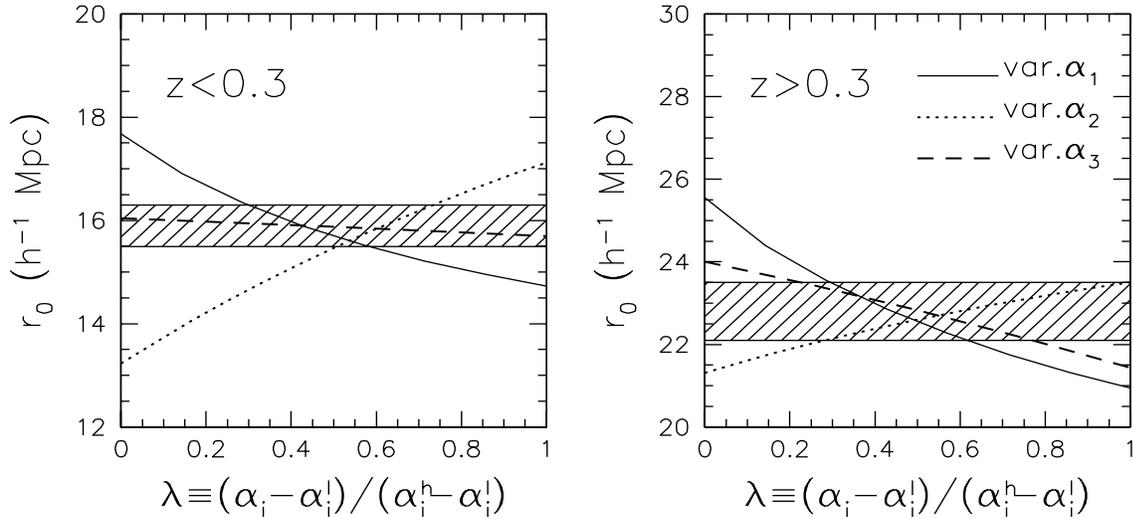,width=16.cm,height=8.cm,angle=0}
\caption{The variation in the predicted correlation length
$r_0$ of galaxy clusters with $z<0.3$ (left panel) and $z>0.3$ (right
panel) is shown in response to changes of the parameters $\alpha_i$ in
the mass-temperature relation. The abscissa range is normalised using
the parameter $\lambda$ defined as $\lambda\equiv
(\alpha_i-\alpha_i^l)/(\alpha_i^h-\alpha_i^l)$, where $\alpha_i^l$ and
$\alpha_i^h$ are the lower and upper limits for the parameter
$\alpha_i$. The parameter $\alpha_1$ is the normalisation, ranging
between 3 and 10 keV; $\alpha_2$ is the exponent of the mass, ranging
between 0.3 and 0.9; and $\alpha_3$ is the exponent of the redshift
dependence, ranging between -1 and 1. Results are presented for the
$\Lambda CDM$ model only. The dashed regions show the 1-$\sigma$
ranges for $r_0$ as obtained using the parameters from
Tab.~\ref{t:models}.}
\label{fi:r0_lambda}
\end{figure*}
%--------------------------------------------------------
%

Figure \ref{fi:r0_lambda} shows the variation of the correlation
length when only one parameter is changed. Results for clusters with
$z<0.3$ and $z>0.3$ are presented in the left and right panels,
respectively. For showing all curves in the same plots, we normalise
the abscissa range using a parameter $\lambda$ defined as
$\lambda\equiv (\alpha_i-\alpha_i^l)/(\alpha_i^h-\alpha_i^l)$, where
$\alpha_i^l$ and $\alpha_i^h$ are the lower and upper limits of the
considered range for the parameter $\alpha_i$. We consider the
following intervals which are expected to cover the whole parameter
region suggested by theoretical and numerical models: $3\le
\alpha_{_1}/{\rm keV} \le 10$, $0.3\le \alpha_{_2} \le 0.9$, $-1\le
\alpha_{_3} \le 1$. As expected, the redshift dependence ($\alpha_{_3}$)
is absolutely negligible in the low-redshift sample, but it also
remains very weak for clusters with $z>0.3$. The results are different
when changing $\alpha_{_1}$ and $\alpha_{_2}$. Increasing the
normalisation of the mass-temperature relation from 3 to 10 keV, $r_0$
is decreased from approximately 18 to 14.5$h^{-1}$ Mpc for $z<0.3$ and
from approximately 26 to 21$h^{-1}$ Mpc for $z>0.3$. Similarly, when
the slope $\alpha_{_2}$ changes from 0.3 to 0.9, the correlation
length increases from $\approx 13$ to $\approx 17 h^{-1}$ Mpc for
$z<0.3$ and from $\approx 21$ to $\approx 23.5 h^{-1}$ Mpc for
$z>0.3$. We recall that the predicted 1-$\sigma$ error bars for $r_0$
are 0.4 and 0.7 for the cluster samples at low- and
high-redshift. Thus, using the clustering properties of SZ galaxy
clusters to constrain cosmological parameters is hampered by our
ignorance of the mass-temperature relation.

\begin{figure*}
\centering
\psfig{figure=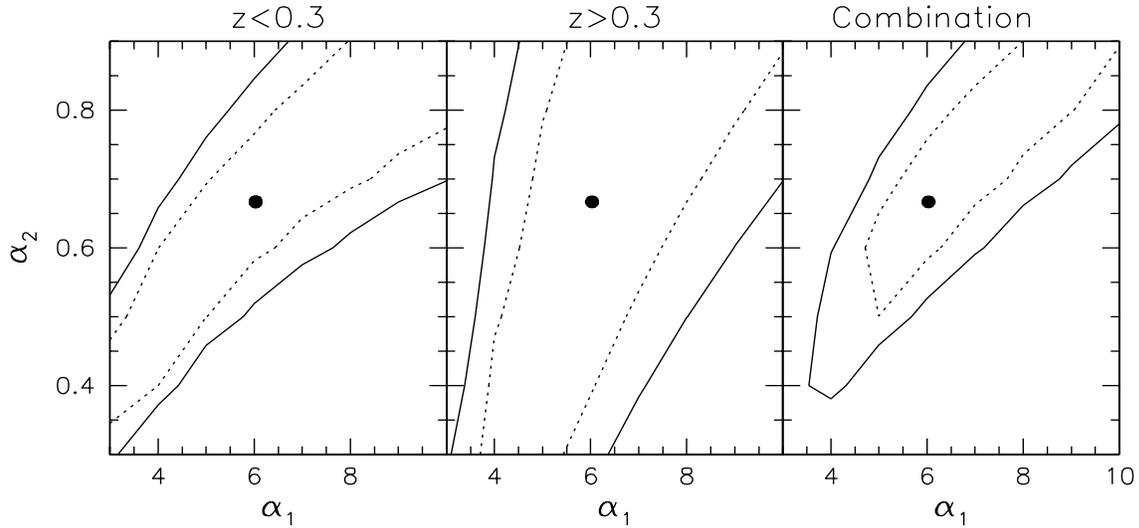,width=16.cm,height=8.cm,angle=0}
\caption{Confidence contours (68.3 and 95.4 per cent confidence levels;
dotted and solid lines, respectively) are displayed for the
normalisation ($\alpha_{_1}$) and slope ($\alpha_{_2}$) of the
mass-temperature relation (see eq. \ref{eq:t-m_mod}); no redshift
evolution is assumed here ($\alpha_{_3}=0$). Results are given for the
$\Lambda$CDM model only. Different panels refer to cluster samples
with $z<0.3$ (left panel), $z>0.3$ (central panel) and to the
combination of the two samples (right panel). Filled circles show the
fiducial values of $\alpha_1=6.03$ and $\alpha_2=2/3$ which are the
input parameters.}
\label{fi:alpha12}
\end{figure*}

However, the dependence of $r_0$ on the details of the
mass-temperature relation can be used to constrain them, once the
cosmological model is known from alternative measurements (for
instance, the analysis of the cosmic microwave background
power-spectrum). To explore this possibility we allow the parameters
to vary and compare with a maximum likelihood analysis the predicted
correlation lengths to the results obtained using
$(\alpha_{_1},\alpha_{_2},\alpha_{_3})=(6.03,2/3,0)$, which is our
input parameter set. We recall that the corresponding correlation
lengths are $r_0=15.9\pm 0.4$ and $22.8\pm 0.7 h^{-1}$ Mpc for
clusters with $z<0.3$ and $z>0.3$, respectively. Since the previous
results showed that the dependence on the redshift evolution parameter
$\alpha_{_3}$ is negligible, we decide to exclude it from the present
analysis. Figure \ref{fi:alpha12} shows the 68.3 and 95.4 per cent
confidence levels for the samples of clusters with $z<0.3$ (right
panel), $z>0.3$ (central panel) and for the combination of the two
samples (right panel). Both at low and high redshift the normalisation
and the slope of the mass-temperature relation appear to be strongly
correlated, allowing large regions in the parameter space. This is
expected: the same minimum mass can be obtained using different
combinations of $\alpha_{_1}$ and $\alpha_{_2}$. When we combine the
results of the two samples, the contours shrink and the parameters are
better constrained. However the degeneracy between the two parameters
does not permit their determination with (68.3 per cent) uncertainties
less than 20 per cent.

\section{Conclusions}

In this paper we presented predictions for the clustering properties
of galaxy clusters detectable through the thermal Sunyaev-Zel'dovich
effect. In particular we showed results for surveys which are expected
from the future {\em Planck\/} satellite, which will cover the whole
sky down to a sensitivity limit of $\overline{Y}_{\rm min}=3\times
10^{-4}$ arcmin$^2$.  Although our estimate of the sensitivity limit
is quite conservative and very likely clusters of smaller mass or at
higher redshift will be included into the real catalogue, the build up
of a reliable cluster sample based on {\em Planck\/} data is not a
straightforward process and requires large efforts, starting from the
simultaneous detection at (at least) two distinct frequencies of the
enhancement and the decrement of the SZ signal and ending to the
source identification and redshift determination.  Despite these
difficulties, it is worth stressing that {\em Planck\/} will give us a
unique chance to construct an independent and unbiased galaxy cluster
sample and to test cosmological scenarios.  Here we showed, in fact,
how it is possible to put independent constraints on some cosmological
parameters and/or on physical properties of the intergalactic medium,
on the basis of such a catalogue.

Our theoretical predictions were obtained for different cosmological
scenarios using a model which accounts for the clustering of
observable objects on our past light-cone and for the redshift
evolution of both the underlying dark matter covariance function and
the cluster bias factor. A linear treatment of redshift-space
distortions was also included.  Following an approach already applied
to X-ray selected clusters, we make use of a theoretical relation
between mass and temperature to convert the limiting sensitivity of a
catalogue into the corresponding minimum mass for the dark matter
haloes hosting the clusters. Based on this relation, the estimates for
the clustering properties allow tests of the general scheme for the
biased formation of galaxy clusters.

We found that the correlation length is an increasing function of the
limits defining the surveys. A similar result was obtained in previous
analyses of optical and X-ray selected clusters. When
$\overline{Y}_{\rm min}$ is fixed to $3\times 10^{-4}$ arcmin$^2$ as
expected for {\em Planck\/}, the large number of detectable objects
will allow a very accurate determination of the correlation length,
with 1-$\sigma$ error bars of few per cent, much better than the
existing estimates. Moreover the {\em Planck\/} cluster sample will
extend to $z\ga 1$, giving the possibility to measure the redshift
evolution of the correlation function. We found that the model
predictions are depending on the main cosmological parameters, like
the matter density parameter $\Omega_{\rm 0m}$ and the power-spectrum
normalisation $\sigma_8$, while the possible effect of the presence of
a cosmological constant is almost negligible. However, the possibility
of using future clustering data obtained by {\em Planck\/} to
constrain the cosmological parameters is limited by the dependence of
the results on the model adopted for the intracluster medium. For
example, different assumptions on the baryon fraction produce
variations of the correlation length similar to those found by
changing $\Omega_{\rm 0m}$. However, the dependence on $\sigma_8$ is
strong enough for it to be constrained, in particular with
high-redshift clusters (see Figure \ref{fi:r0_dep_new}).

When a set of cosmological parameters is fixed, the cluster
correlation function depends on the properties of the intracluster
medium. We found that the {\em Planck\/} data can be used to constrain
the baryon fraction with an accuracy of few per cent once the value of
the Hubble constant is known. On the contrary, our results showed that
the relation between mass and temperature, which parameterises the
whole history of the physical processes inside the clusters, can only
be poorly determined from clustering data alone.

\section*{Acknowledgments.}

This work has been partially supported by Italian MIUR (Grant 2001,
prot. 2001028932, ``Clusters and groups of galaxies: the interplay of
dark and baryonic matter''), CNR and ASI.  We are grateful to
Cristiano Porciani and Bepi Tormen for clarifying
discussions. P.A. warmly thanks the IR group of the Max-Planck
Institut fuer Extraterrestrische Physik in Garching for hospitality.

\end{document}